\documentclass[11pt]{article}
\usepackage{moriond,epsfig}

\def\mco{\multicolumn}

\def\be{\begin{equation}}
\def\ee{\end{equation}}
\def\bea{\begin{eqnarray}}
\def\eea{\end{eqnarray}}

\begin{document}
\vspace*{4cm}
\title{DIBOSON PHYSICS AT THE TEVATRON
\footnote{Contribution to the Proceedings of the XLIst
Rencontres de Moriond: Electroweak Interactions and Unified Theories, La Thuile
(Italy), March 11-18, 2006.}
}

\author{ J. Sj\"olin\\(On behalf of the CDF and D0 collaborations)
}

\address{Department of Physics, Keble Road, Oxford OX1 3RH, England.\\
Niels Bohr Institute, Blegdamsvej 17, DK - 2100 Copenhagen.}

\maketitle\abstracts{
Measurements of diboson events in proton-antiproton collisions at
$\sqrt{s}=1.96$ TeV are effective probes of the electro-weak gauge structure
in the
Standard Model (SM). Due to the potentially high energy scale in the
collisions, bosonic operators beyond the leading order
in the Lagrangian become important at the Tevatron and make the theoretical
interpretations non trivial. The fact that we need cut-offs in our models
in order to maintain unitarity underline the importance of
direct measurements at the highest possible energy scales.
I report on two new such measurement from CDF and compare to similar D0
results taken at lower integrated luminosities.
}

\section{Introduction to dibosons}
Processes involving multiple massive bosons probe right at the
core of the SM, i.e. its electro-weak symmetries and resulting
gauge boson structure. Our current understanding requires some -- yet
undetected --
mechanism to regularize processes like $W_L W_L \rightarrow W_L W_L$ at the
$O(1)$ TeV scale in order to maintain unitarity \cite{Chanowitz:2004gk}.
This hints at an experimental no-lose situation where we either will
find the scalar
responsible for the Higgs mechanism, or find new interactions
interacting strongly with the longitudinal gauge modes. If none of these cases
are seen we will be obliged to go back to the sub TeV region and ask ourself
what it is that we don't understand on the fundamental level.

Our lack of knowledge concerning the underlying electro-weak symmetry
breaking (EWSB) mechanism of the SM desperately calls for experimental input to
guide us. Clearly, as indicated from the discussion above diboson
scattering is likely to provide the ultimate EWSB exploration tool.
Unfortunately the cross-section for diboson scattering is expected to
be unobservable at the Tevatron compared to the LHC \cite{Butterworth:2002tt}.
Another way to get a handle on the EWSB is to reconstruct massive triple gauge
bosons, but also here is the expected Tevatron cross-section beyond reach.
It is primarily for inclusive dibosons that we currently are gaining
sensitivity to anomalous gauge interactions where new phase-space regions
are opening up with the increase of integrated Tevatron luminosity.

LEP 2 puts stringent limits on
the anomalous triple gauge couplings \cite{Bruneliere:2004ab} (ATGC) which
are the lowest order general effective theory operator couplings related
to the diboson final state. However, the LEP 2 energy scale
$\sqrt{\hat{s}} \sim 200$ GeV is sufficiently small compared to the new
physics scale $\Lambda$ such that higher order operators can be neglected. At
the Tevatron this is no longer the case since $O(1)$ TeV resonances can
be produced on-shell and a form factor must be
introduced in order to maintain unitarity of the model \cite{Hagiwara:1989mx}.
The form factor is equivalent to a infinite series of higher order operators
and implies a non linear extension of the linear model commonly used
at LEP. For this reason the quoted limits on the couplings refer to
slightly different assumptions at the two accelerators. And one should be
aware of that Tevatron measurements are in many ways complementary to
those made at LEP, e.g. new unexpected
on-shell dynamics may kick-in at higher energies and with different
production mechanisms.

\section{Tevatron measurements}
As can be seen from table \ref{tab:diboxs} it is just recently that
massive diboson signals became statistically significant at hadron colliders.
\begin{table}[t]
\caption{New(**) and recent diboson measurements at the Tevatron. The
photon($\gamma$) has a transverse momentum cut of 7(8) GeV for the
CDF(D0) measurements. All cross-sections are within the theoretical
expectations.
\label{tab:diboxs}}
\vspace{0.4cm}
\begin{center}
\begin{tabular}{|c|c|c|c|c|}
\hline
& \mco{2}{|c|}{CDF} & \mco{2}{|c|}{D0} \\
\hline
Diboson & Events & $\sigma$[pb] & Events & $\sigma$[pb] \\
\hline
$W(l\nu)\gamma \times BR(l\nu)$ & 208(200/pb)& $18.1 \pm 3.1$ &
141(162/pb) & $14.8 \pm 1.9$ \\
$Z(ll)\gamma \times BR(ll)$ & 66 (200/pb)& $4.6 \pm 0.6$ &
244(320/pb) & $4.2 \pm 0.5$ \\
$W(l\nu)W(l\nu)$ & 12(200/pb) &
\begin{minipage}{1in}
$14.6^{+5.8}_{-5.1}$(stat) $^{+1.8}_{-3.0}$(sys) $\pm 0.9$(lum)
\end{minipage}
& 17(252/pb) &
\begin{minipage}{1in}
$13.8^{+4.3}_{-3.8}$(stat) $^{+1.2}_{-0.9}$(sys) $\pm 0.9$(lum)
\end{minipage}
\\
$W(l\nu)(W(jj)+Z(jj))$ & n.a.(350/pb) & $<36$ (**) & & \\
$W(l\nu)Z(ll)$ & n.a.(825/pb)& $<6.4$ (**) & n.a.(320/pb) & $<13.3$ \\
\hline
\end{tabular}
\end{center}
\end{table}
The measured cross-sections decrease with the diboson mass and the first
process that has been firmly established during the last two years is WW to
leptons. This is very different from LEP 2 where thousands of WW pairs have
been observed in many decay channels, both leptonic and pure hadronic
modes. However, constraints on new physics are dominated by high
energy scales and the inclusive cross-section is not the most efficient
observable at the Tevatron for anomalous gauge interactions. A closer
look at the anomalous cross-section contributions \cite{Hagiwara:1989mx}
reveal that they, apart from a boson angular dependence, scale with
the parton center-of-mass energy ($\sqrt{\hat{s}}$). Thus the very high energy
tail of the boson transverse momentum turns out to be a extremely sensitive and
robust observable for ATGC.

In the following subsections I will report on two brand new measurements
of WW and WZ dibosons from CDF. It is interesting to note that even though the
processes themselves are just below the the threshold of observation they
have a significant constraining power for ATGC.

\subsection{The process $WZ \rightarrow l\nu ll$}

\begin{figure}
\begin{center}
\psfig{figure=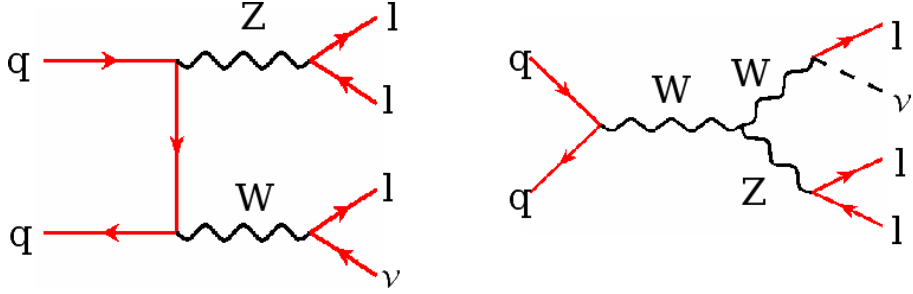,width=12cm}
\caption{Tree level graphs for WZ.
\label{fig:wz}}
\end{center}
\end{figure}

The WZ process, see figure \ref{fig:wz}, is particularly interesting among
the dibosons.
Since it could not be be produced at tree level at LEP we have so far no
direct experimental measurement of the WWZ vertex. Furthermore,
the absence of interference with other triple gauge vertices at tree level
makes this channel unique since it can provide a unambiguous
handle on the WWZ couplings in case of any observed anomaly that needs to be
disentangled in e.g. the WW production.

\begin{figure}[p]
\begin{center}
\psfig{figure=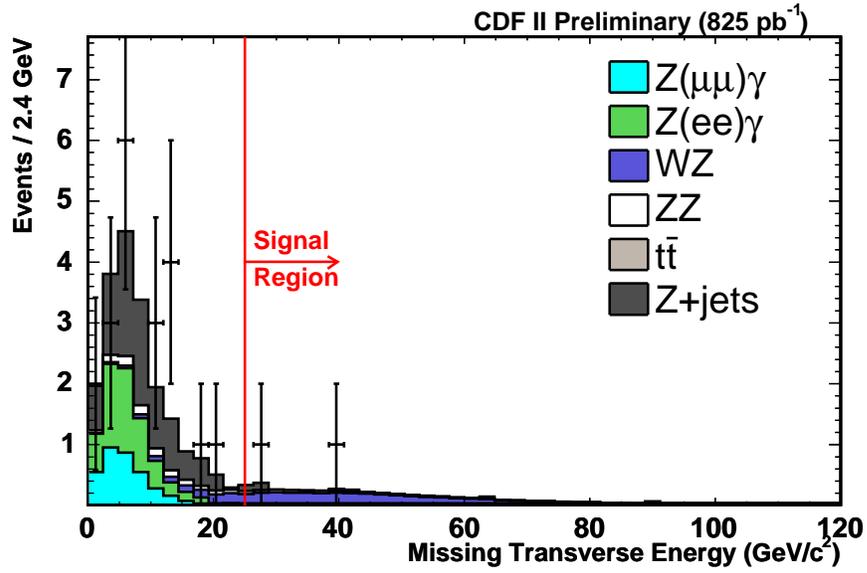,width=13cm}
\caption{Missing transverse energy in the WZ events. The cut for the signal
region is also shown.
\label{fig:wzmet}}
\end{center}
\end{figure}

\begin{figure}[p]
\begin{center}
\psfig{figure=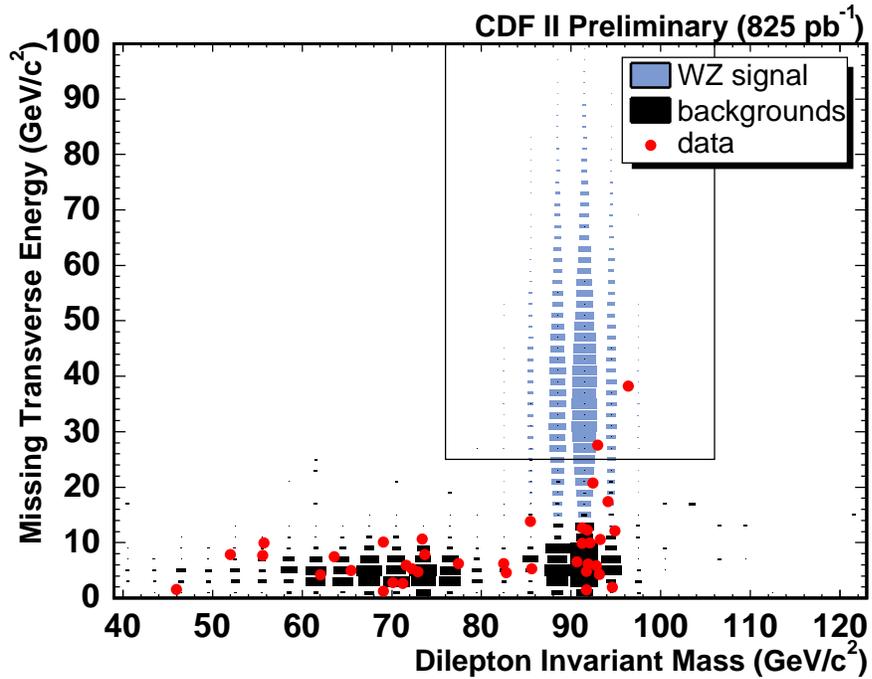,width=13cm}
\caption{Scatter plot in the missing transverse energy and dilepton mass
plane for WZ. Data is compared to expectation.
\label{fig:wzmll}}
\end{center}
\end{figure}

A new CDF search for WZ to leptons has been made using data
equivalent to 825/pb of integrated luminosity of proton-antiproton collisions
at $\sqrt{s}=1.96$ TeV. The events are triggered by a lepton (isolated
electron or muon) with at least 20 GeV transverse momentum. Two more leptons
are then required with at least 10 GeV transverse momentum. A neutrino
like signature is selected by requiring at least 25 GeV, see figure
\ref{fig:wzmet}, of missing transverse
energy. One Z is selected by one opposite sign lepton pair within the
dilepton mass $76 < M_{ll} < 106$ GeV, see figure \ref{fig:wzmll},
and ZZ events are vetoed by removing
events with tracks that falls within $76 < M_{trk,l} < 106$ GeV using the
remaining unmatched lepton. Due to the specific multilepton signature the
selected candidates have a relatively small expected background contamination.
An overview of measured and estimated events are shown in
table \ref{tab:wz}. Only two candidates are observed and a limit is
derived on the total cross-section
$$
\sigma(p\bar{p} \rightarrow WZ)< 6.4 \textrm{ pb (95\% C.L.)}.
$$
This is intriguingly close to the theoretical NLO value around 4 pb.
Limits on the WWZ ATGC have not yet been extracted but we know
from a recent and similar D0 measurement \cite{Abazov:2005ys}
(see also the corresponding entry in table 1)
that this analysis have the potential to provide strong constraints.

\begin{table}[t]
\caption{Measured and expected number of events for WZ to leptons
at CDF using 825/pb of data.
\label{tab:wz}}
\vspace{0.4cm}
\begin{center}
\begin{tabular}{|l|l|}
\hline
Process & Events \\
\hline
$WZ$ & $3.72 \pm 0.02$(stat) $\pm 0.15$(sys) \\
\hline
$ZZ$ &  $0.50 \pm 0.01$(stat) $\pm 0.05$(sys) \\
$Z\gamma$ &  $0.03 \pm 0.01$(stat) $\pm 0.01$(sys) \\
$t\bar{t}$ &  $0.05 \pm 0.01$(stat) $\pm 0.01$(sys) \\
$Z+jets$ &  $0.34 \pm 0.07$(stat) $^{+0.16}_{-0.10}$(sys) \\
\hline
Total background: &  $0.92 \pm 0.07$(stat) $\pm 0.15$(sys) \\
\hline
Total data: &  2 \\
\hline
\end{tabular}
\end{center}
\end{table}

\subsection{The process $WW+WZ \rightarrow l\nu jj$}

\begin{figure}
\begin{center}
\psfig{figure=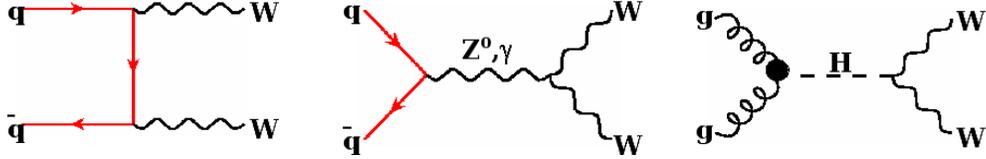,width=13cm}
\caption{Tree level and LO Higgs graphs for WW.
\label{fig:ww}}
\end{center}
\end{figure}

Semi-leptonic decay modes of WW events have yet not been observed at hadron
colliders, see figure \ref{fig:ww} where one of the bosons decays into quarks
and the other one into leptons.
The reconstruction is challenging due
to the limited dijet mass resolution of about 10\% and the W+jets background
which is $O(100)$ times larger than the signal after event preselection
at the Tevatron. Note that events from diagrams in figure \ref{fig:wz} (replace
the leptonic Z decay with quarks) cannot be excluded, hence they are included
in the signal. Still, the hadronic decay modes are interesting since
they have larger event yield than the pure leptonic modes and they have
the important ability to reconstruct the W transverse momentum which we
know is a sensitive handle on the ATGC.

In order to constrain the ATGC and to verify the SM rate
a new CDF search for WW+WZ to leptons, missing transverse energy and jets
has been made using data equivalent to 350/pb of integrated luminosity
of proton-antiproton collisions at $\sqrt{s}=1.96$ TeV. The events are
triggered by an isolated lepton with 25 GeV transverse energy (20 GeV
transverse momentum for muons). An inclusive set of leptonic W decays is
selected by requiring the missing transverse energy to be at least 25 GeV.
The events selection then proceeds by requiring two jets with at least
15 GeV transverse energy and with a dijet mass in the range $32 < M_{jj} <
184$ GeV. An additional cut on the W transverse mass below 25 GeV is applied
to reduce the multijet background. Also proximity cuts are applied among the
lepton and jets. However care must be taken not to reject narrow jets since
the interesting high transverse momentum boson have narrow jets due to the
boost. The expected number of events is shown in table \ref{tab:ww}.

\begin{table}[t]
\caption{Measured and expected number of events for WW+WZ to leptons,
missing transverse energy and jets at CDF using 350/pb of data.
\label{tab:ww}}
\vspace{0.4cm}
\begin{center}
\begin{tabular}{|l|c|r|}
\hline
Process & Uncertainty & Events \\
\hline
$WW$ & 15\% & 142.0 \\
$WZ$ & 15\% & 18.2 \\
\hline
$W+jets$ & 20\% & 6261.0 \\
$Multijets$ & 40\% & 263.4 \\
$W(\tau)+jets$ & 20\% & 171.0 \\
$Z+jets$ & 20\% & 154.0 \\
$t\bar{t}$ & 25\% & 171.6 \\
$t(t-ch)$ & 25\% & 14.4 \\
$t(s-ch)$ & 25\% & 8.2 \\
\hline
\end{tabular}
\end{center}
\end{table}

\begin{figure}[t]
\begin{center}
\psfig{figure=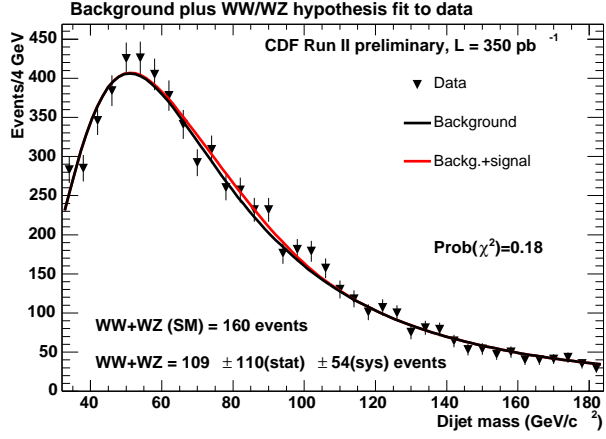,width=9cm}
\caption{WW+WZ signal and background dijet mass spectrum shape fitted to the
measured data. This is used for estimating the total cross-section.
\label{fig:wwxsfit}}
\end{center}
\end{figure}

Since the theoretical uncertainty on the W+jets background is huge we extract
the WW+WZ signal by assuming the SM ratio between WW/WZ and fit the expected
signal and background dijet mass shapes to data, see figure \ref{fig:wwxsfit}.
The fit gives 109 signal events and an statistical uncertainty of 110 events.
Systematic uncertainties contributes with an additional 54 events. From this
we estimate the upper limit on the WW+WZ cross-section to be
$$
\sigma(p\bar{p} \rightarrow WW+WZ) < 36 \textrm{ pb}.
$$
The SM expectation is 16 pb.

\begin{figure}[t]
\begin{center}
\psfig{figure=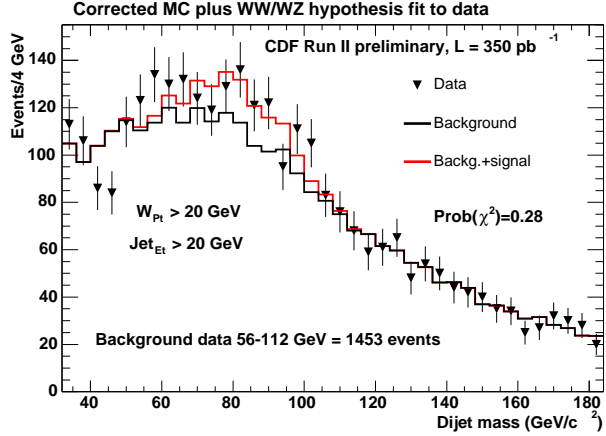,width=9cm}
\caption{WW+WZ signal and background dijet mass spectrum shape fit
with cuts used for estimating the ATGC background.
\label{fig:wwbg}}
\end{center}
\end{figure}

\begin{figure}[t]
\begin{center}
\psfig{figure=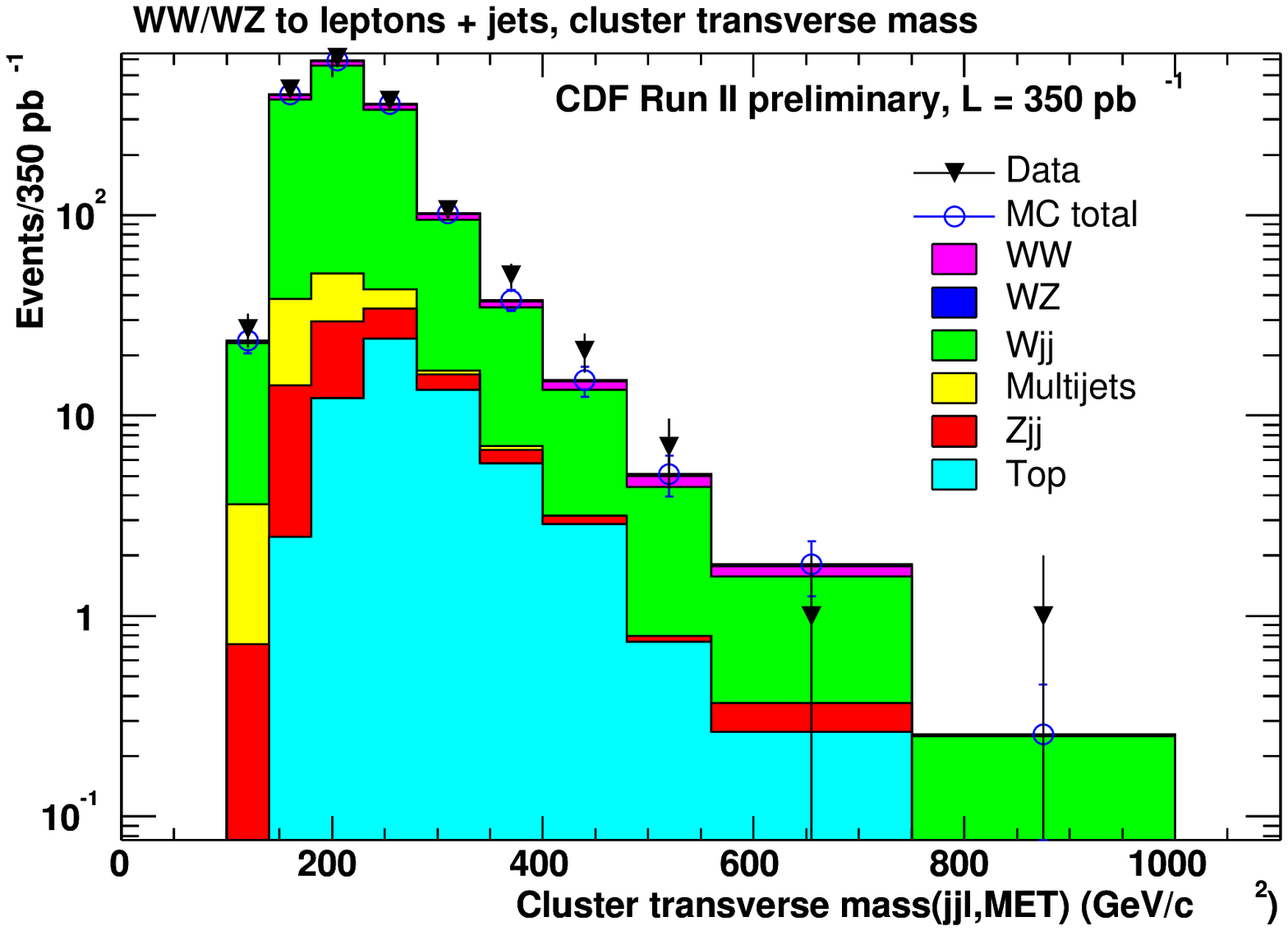,width=11cm}
\psfig{figure=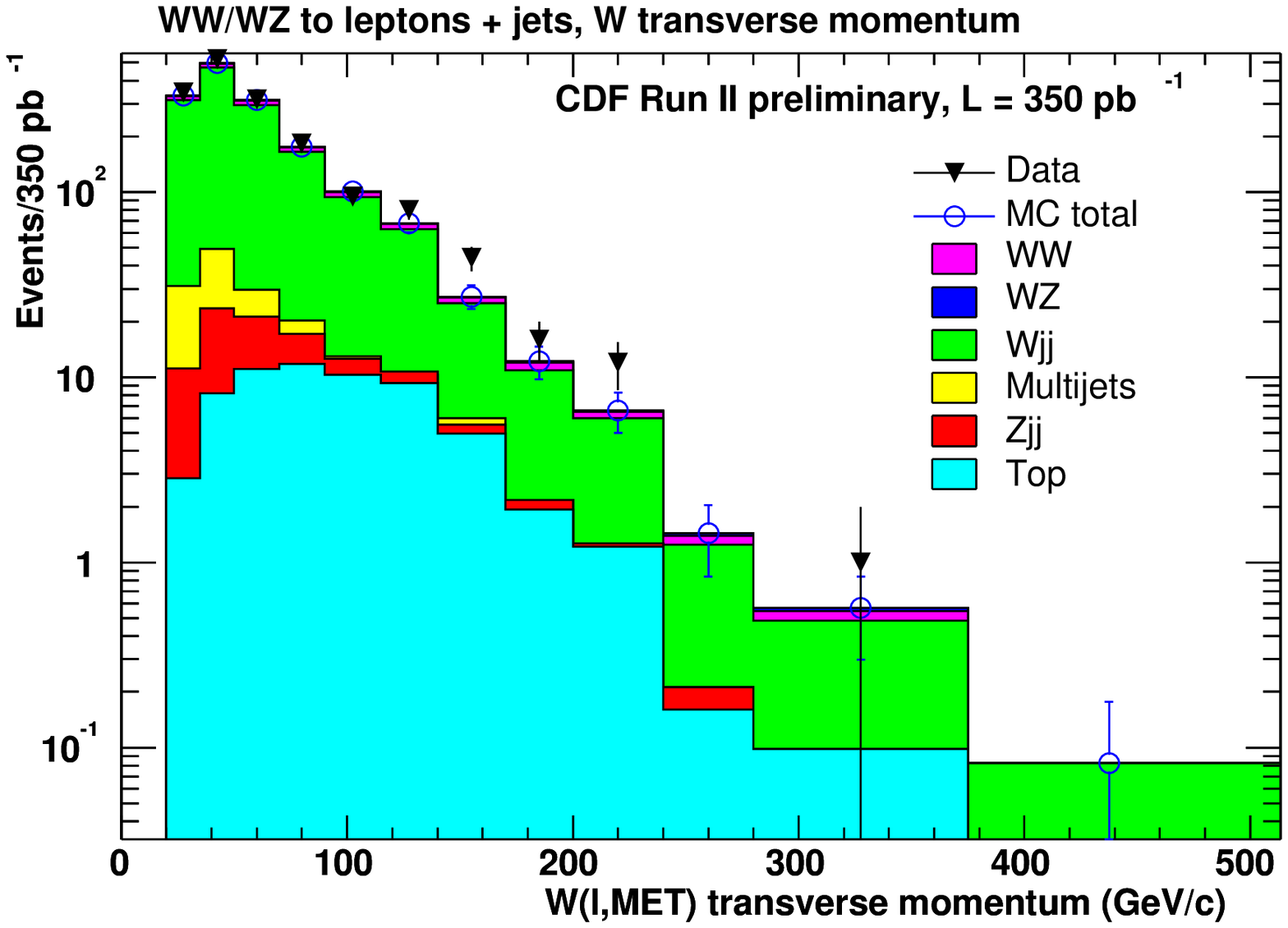,width=11cm}
\caption{Cluster transverse mass (upper plot) and W transerse momentum
(lower plot).
\label{fig:wwpt}}
\end{center}
\end{figure}

\begin{figure}[t]
\begin{center}
\psfig{figure=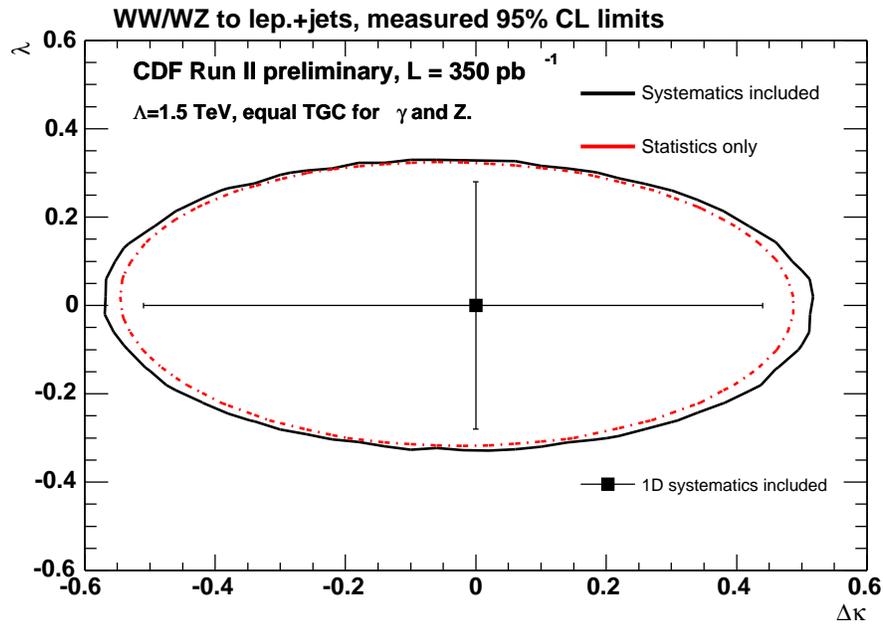,width=13cm}
\caption{ATGC limits extracted from the W transverse momentum in
WW+WZ to lepton + jets events. The 1D limit shows the case when one
of the parameters is fixed to zero.
\label{fig:ac}}
\end{center}
\end{figure}

The ATGC are extracted from the signal region defined to be the
dijet mass window $56<m_{jj}<112$ GeV. The fit and extra cuts used for
estimating the ATGC background events are shown in figure \ref{fig:wwbg}.
The actual ATGC limits are given by a binned likelihood fit to the
W transverse momentum, see figure \ref{fig:wwpt}. For completeness we also show
the transverse cluster mass which is a good estimator of $\hat{s}$ in the event
and a probe for unexpected resonances. Finally the 2D ATGC limits in the
$(\Delta\kappa,\lambda)$ plane is shown in figure \ref{fig:ac}. The
corresponding 1D limits where one of the parameters are fixed to zero are
$$
-0.51 < \Delta\kappa < 0.44
$$
$$
-0.28 < \lambda < 0,28
$$
This is with a form factor cut-off $\Lambda=1.5$ TeV. Previous limits in a
comparable channel are from D0 and Run I \cite{Abbott:1999aj}.
The quoted limits are
$$
-0.56 < \Delta\kappa < 0.75
$$
$$
-0.42 < \lambda < 0.44
$$

\section*{Acknowledgments}
I want to thank the Moriond organizers for providing
an exceptionally stimulating conference environment.
Also a big thanks to my colleagues and collaborators at the D0 and CDF
experiments.

\end{document}